\documentclass[reprint,aps,prb,twocolumn,longbibliography,superscriptaddress,nobalancelastpage]{revtex4-2} 

\usepackage{graphicx}   
\usepackage{amsmath,amssymb}
\usepackage[colorlinks=true, allcolors=blue]{hyperref}

\usepackage{physics}
\usepackage{xcolor}
\usepackage{soul}
\usepackage{comment}
\usepackage[normalem]{ulem}
\usepackage{enumitem}
\usepackage[T1]{fontenc}

\newcommand{\be}{\begin{equation}}
\newcommand{\ee}{\end{equation}}
\newcommand{\bea}{\begin{eqnarray}}
\newcommand{\eea}{\end{eqnarray}}
\newcommand{\rr}{\boldsymbol{\rm r}}
\newcommand{\rrp}{\boldsymbol{\rm r}^\prime}

\newcommand{\aae}[1]{\textcolor{black}{#1}}
\DeclareFontFamily{OT1}{pzc}{}
\DeclareFontShape{OT1}{pzc}{m}{it}%
              {<-> s * [1.180] pzcmi7t}{}
\DeclareMathAlphabet{\mathpzc}{OT1}{pzc}%
                                 {m}{it}
\begin{document}
\title{ First-principles simulation of shocked H-He mixture along the principal Hugoniot}
\author{Ammar A.~Ellaboudy}
\affiliation{Laboratory for Laser Energetics,
University of Rochester, 250 East River Road, Rochester NY 14623, USA}
\affiliation{Department of Physics and Astronomy, University of Rochester, Rochester, NY 14627, USA}
\author{Valentin V.~Karasiev}
\email{vkarasev@lle.rochester.edu}
\affiliation{Laboratory for Laser Energetics,
University of Rochester, 250 East River Road, Rochester NY 14623, USA}
\author{S. X. Hu}
\affiliation{Laboratory for Laser Energetics,
University of Rochester, 250 East River Road, Rochester NY 14623, USA}
\affiliation{Department of Physics and Astronomy, University of Rochester, Rochester, NY 14627, USA}
\affiliation{Department of Mechanical Engineering, University of Rochester, Rochester, NY 14627, USA}

\date{02 May 2026} 

\begin{abstract}

Recent laser-shock experiments on an H--He mixture containing 11~$\%$ helium (atomic fraction) have suggested the presence of an immiscibility region inside Jupiter. Reflectivity measurements were used as the primary diagnostic of H--He demixing, with discontinuities in the optical reflectivity proposed as a signature of phase separation under conditions relevant to Jupiter's interior. Here, we investigate shock-compressed H--He using \textit{ab initio} molecular dynamics simulations with optical properties evaluated within the Kubo--Greenwood formalism. The equation of state and ionic configurations were obtained using the thermal Tr$^2$SCANL meta-GGA exchange--correlation (XC) functional, while optical properties were computed using the recently developed RS-KDT0 range-separated thermal hybrid XC, which provides state-of-the-art accuracy for band-gap predictions in the warm dense matter regime. The calculated reflectivity shows overall good agreement with experimental measurements; however, no discontinuity is observed at elevated temperatures. 
Moreover, the reflectivity predictions for the mixed system are consistent with the experimental measurements in the temperature range where the mixture is inferred to be demixed. These results suggest that reflectivity alone may not provide a unique or sensitive diagnostic of H-He demixing at low helium concentrations under these conditions.

\end{abstract}

\maketitle





\section{Introduction}

The properties of hydrogen--helium (H-He) mixtures in the warm dense matter regime are central to modeling the interiors of giant gas planets such as Jupiter and Saturn. Current models of giant planets~\cite{Wahl_comparing_Jupiter_interior_2017,Debras_2019,Mankovich_2020,Howard2024HeliumRain,Sur_2025,Tejada_Arevalo_2024} predict a layered internal structure consisting of an outer homogeneous envelope of molecular hydrogen and helium, followed by a He-enriched layer associated with H-He immiscibility, \aae{a phase-separated mixture in which atoms of the same species aggregate toward one another}.
 Beneath this region lies an inner layer of metallic H-He mixture, and, at greater depths, a gradually eroding core in which the concentration of heavy elements increases with pressure. The existence of an immiscible H-He layer has been proposed to explain both the depletion of atmospheric helium and Saturn’s excess luminosity~\cite{Stevenson_1975}. Accurately determining the location and thickness of these layers requires precise knowledge of the H-He immiscibility boundaries in temperature--pressure space for a given helium concentration.

The immiscibility properties of H–He mixtures have been investigated theoretically using density functional theory (DFT) by calculating the Gibbs free energy of mixing. From the Gibbs free energy, the critical properties—including the demixing temperature, and pressure for each helium fraction—are obtained using the double-tangent construction~\cite{Lorenzen_H-He_2009,Morales_H-He_2013,schottler_H-He_2018}. However, computing the Gibbs free energy is challenging because the entropy of mixing cannot be accessed directly from DFT simulations. Some studies have estimated the immiscibility boundary by employing a linear mixing approximation for the entropy~\cite{Lorenzen_H-He_2009,schottler_H-He_2018}. More accurate DFT studies using PBE~\cite{Morales_H-He_2013} and van der Waals~\cite{schottler_H-He_2018} exchange–correlation (XC) functionals have instead relied on coupling constant integration, and thermodynamic integration, to evaluate the entropy contribution. Quantum Monte Carlo benchmarking of XC functionals for H--He mixtures~\cite{Raymond_benchmarking_XC_2016} shows that van der Waals functionals outperform PBE in predicting the enthalpy of mixing, which is a key contribution to the Gibbs free energy of mixing. However, PBE yields smaller discrepancies in the calculated mixing temperature compared to van der Waals functionals~\cite{Raymond_benchmarking_XC_2016}: 
the immiscibility boundary predicted by van der Waals XC functional appears to be systematically colder than Jupiter's adiabat.  When intersected with Jupiter’s adiabat, such phase diagrams predict no immiscibility within the planet conflicting with the observed atmospheric helium depletion; consequently, these boundaries are often shifted to higher temperatures to obtain interior models consistent with observations~\cite{Mankovich_2020,Howard2024HeliumRain}. Additionally, the immiscibility boundaries for an 11.3 and 27.5 ~$\%$~(He atomic fraction) H--He mixtures have been predicted using the thermal KDT16 GGA XC functional to account for finite-temperature effects, yielding the boundary location about 500 K higher as compared to PBE predictions.
Demixing was investigated by monitoring changes in the height of the first peak of the H-He radial distribution function (RDF), providing direct structural insights, while avoiding finite size effects and the computational complexity of the Gibbs free-energy approach \cite{karasiev2026_H-He}. Furthermore, H--He mixtures were studied using large-scale simulations based on neural-network potentials (NNPs) trained on \textit{ab initio} molecular dynamics (AIMD) data~\cite{Chang2024HHeDemixing}. In these calculations, demixing was analyzed directly from the simulation data using a nearest-neighbor unweighted conditional probability, yielding demixing temperatures higher than those predicted by AIMD simulations \cite{karasiev2026_H-He}.

Recent laser-driven shock experiments on H$_2$--He mixtures have revealed evidence of demixing along the principal Hugoniot under thermodynamic conditions relevant to Jupiter’s interior~\cite{Byrgoo_H-He_2021}. In these experiments, the shock-induced reflectivity measured by a line-imaging velocity interferometer (VISAR) is used as a signature of demixing, based on the premise that the optical properties of mixed (\aae{homogeneous}) and demixed systems (\aae{phase separated}) at the same temperature and pressure are expected to differ~\cite{Soubrian_Optical_2013}. For a mixture containing 11\% helium atomic fraction, two distinct reflectivity jumps/drops were observed: an increase at approximately 4{,}700~K followed by a decrease near 10{,}200~K. These observations were interpreted as indicator that the mixture enters the immiscibility regime between these two temperatures. By combining these results with earlier static compression experimental measurements at lower pressures~\cite{Loubeyre_H-He_1987} and slop calculated in~\cite{Schouten_H-He_1991}, an immiscibility boundary was constructed that intersects Jupiter’s adiabat, suggesting the presence of a demixing region within the planet. However, the experimentally inferred immiscibility temperature at the boundary around $\sim 150$~GPa is approximately 10{,}200~K, which is about 4{,}000~K higher than predictions from existing theoretical calculations. 

In this work, we perform 
AIMD simulations driven by DFT to investigate the properties of H--He mixtures with $11.3~\%$ He atomic fraction along the principal Hugoniot under conditions relevant to previous laser-driven shock experiments. We qualitatively assess the degree of mixing using H-He RDFs. In addition, we compute the optical properties using the Kubo--Greenwood formalism and directly compare our results with experimental measurements.

The methodology used in this work is described in Sec.~\ref{sec:methods}. 
Section~\ref{sec:comp_details} presents the computational details, 
Sec.~\ref{results} discusses the main results, 
and Sec.~\ref{summary} provides a brief summary and concluding remarks.

\section{Methods}
\label{sec:methods}
In AIMD, the ionic degrees of freedom are treated classically, while the electronic degrees of freedom are described quantum mechanically. Within finite-temperature DFT-based AIMD 
under the Born–Oppenheimer approximation, the electronic density is recalculated at each molecular-dynamics time step using DFT. The free energy of a quantum many-body system is expressed as a functional of the electronic density. The ground state electronic density is obtained by minimizing this free-energy functional, which is achieved by solving the Mermin-Kohn–Sham (MKS) equations \cite{kohn-consistent-equations-xc-1965, mermin_thermal_1965}. The accuracy of DFT critically depends on the approximation used for the exchange-correlation (XC) functional. Extensive effort has therefore been devoted to developing XC approximations of increasing sophistication and accuracy, commonly organized within the framework of Jacob’s ladder~\cite{Jacob-ladder}. A widely used example at the generalized gradient approximation (GGA) level is the Perdew–Burke–Ernzerhof (PBE) functional~\cite{perdew_generalized_1996}.

At elevated temperatures, a common practice is to use the ground-state approximation (GSA), in which the XC functional has no explicit temperature dependence and thermal effects enter only implicitly through the electronic density. However, in the warm dense matter regime, thermal effects become significant~\cite{karasiev_importance_2016}, necessitating the use of finite-temperature XC functionals with explicit temperature dependence. Several such thermal functionals have been developed, including the local density approximation (LDA) level corrKSDT~\cite{karasiev_accurate_2014}, the GGA-level KDT16~\cite{karasiev_nonempirical_2018}, the meta-GGA-level Tr$^2$SCANL, TSCANL~\cite{Karasiev_meta-GGA_2022} and $f$TSCAN~\cite{Katie_fTSCANL_2025}, and the hybrid functionals KDT0~\cite{mihaylov_thermal_2020} and RS-KDT0~\cite{Ammar_RSKDT0_2025}. 

The choice of XC approximation for a given problem requires balancing accuracy against computational cost. For molecular dynamics simulations, meta-GGA XC functionals provide an optimal compromise, offering high accuracy while remaining computationally feasible for thousands of MD steps. In contrast, hybrid XC functionals yield more accurate band-gap predictions~\cite{Ammar_RSKDT0_2025} but are prohibitively expensive for large-scale MD simulations, making them better suited for calculations of transport and optical prope`rties, where only a limited number of representative MD snapshots are required.

\subsection{Thermal meta-GGA XC}

Meta-GGA XC functionals generally provide higher accuracy than GGA- and LDA-level approximations by incorporating additional information about the electronic density. While GGA XC functionals depend on the density and its gradient, meta-GGA functionals further include higher-order ingredients such as the density Laplacian or the kinetic-energy density, enabling a more accurate description of inhomogeneous electronic systems.

Among the most successful ground-state meta-GGA XC functionals are SCAN and its regularized-restored variant r$^2$SCAN~\cite{SCAN_2015,r2SCAN_2019}, which satisfy an extensive set of exact constraints while maintaining broad accuracy across diverse bonding environments. Meta-GGA XC functionals depend explicitly on the Kohn--Sham orbitals through the kinetic-energy density. Deorbitalized variants have been developed in which the orbital dependence is replaced by dependence on the reduced density Laplacian, $q = \nabla^2 n/4 (3\pi^2)^{2/3} n^{5/3}$. The deorbitalized counterparts of SCAN and r$^2$SCAN are denoted SCANL and r$^2$SCANL, respectively. Finite-temperature extensions of meta-GGA XC functionals have been developed by partially thermalizing their ground-state counterparts. In particular, the thermal meta-GGA XC functional Tr$^2$SCANL is constructed by augmenting r$^2$SCANL with a GGA-level thermal correction. This correction is defined using the finite-temperature GGA functional KDT16 and its zero-temperature limit, PBE,
\be
\Delta {\cal F}_{\mathrm{xc}}^{\mathrm{GGA}}[n,T]=
{\cal F}_{\mathrm{xc}}^{\mathrm{KDT16}}[n,T]-E_{\mathrm{xc}}^{\mathrm{PBE}}[n]
\,,
\label{DeltaFxc}
\ee
and added to the ground-state meta-GGA functional according to
\be
    {\cal F}_{\mathrm{xc}}^{\mathrm{meta\text{-}GGA}}[n,T]=
E_{\mathrm{xc}}^{\mathrm{meta\text{-}GGA}}[n]
+ \Delta {\cal F}_{\mathrm{xc}}^{\mathrm{GGA}}[n,T]
\,.
\label{metaGGA}
\ee
For Tr$^2$SCANL, the thermal correction is applied to r$^2$SCANL, while the partially thermalized SCANL (TSCANL) functional is constructed in an analogous manner. Fully thermalized orbital-dependent meta-GGA XC functionals have also been developed by explicitly incorporating temperature dependence into the iso-orbital indicator $\alpha(n,\nabla n,\psi)$ and GGA-level reduced density gradients. Further discussion is given in Ref.~\cite{Katie_fTSCANL_2025}.

\subsection{Range-separated thermal hybrid XC}

A fundamental limitation of MKS-DFT with semi-local XC functionals 
is the systematic underestimation of electronic band gaps. This deficiency arises because the Kohn--Sham band gap, defined as the difference between the lowest unoccupied and highest occupied KS eigenvalues, lacks the derivative discontinuity present in the true fundamental gap. The generalized Kohn--Sham formalism partially remedies this issue by allowing for non-local potentials~\cite{Perdew_understanding_band_gap_2017,Mori-Sanchez_delocalization_error_2008,Cohen_challenges_DFT_2012}.

Hybrid XC functionals exploit this framework by mixing a fraction of non-local exact Fock exchange with a semi-local density functional approximation. Prominent ground-state examples include the PBE0 global hybrid and the HSE range-separated (RS) hybrid~\cite{Perdew_rationale_1996,heyd_hybrid_2003}. Finite-temperature hybrid XC functionals are obtained by thermalizing these ground-state hybrids, leading to the global hybrid KDT0 and the range-separated hybrid RS-KDT0~\cite{mihaylov_thermal_2020,Ammar_RSKDT0_2025}.

In the KDT0 global hybrid functional, a fraction of the thermal Fock exchange is mixed with the KDT16 GGA-level exchange, while the correlation contribution is entirely described by KDT16. Although global hybrids significantly improve band-gap predictions relative to GGA and LDA functionals, they tend to overestimate band gaps in semiconductors due to the absence of electronic screening in the long-range Fock exchange.

Range-separated hybrids address this limitation by partitioning the Coulomb kernel into short-range (SR) and long-range (LR) components using the error function,
\be
\frac{1}{\vert \rr - \rrp \vert}=
\frac{\text{erfc}(\mu\vert \rr - \rrp \vert)}{\vert \rr - \rrp \vert}
+
\frac{\text{erf}(\mu\vert \rr - \rrp \vert )}{\vert \rr - \rrp \vert}
\;,
\label{coulombspliiting}
\ee
where $\mu$ is the Coulomb screening parameter. The RS-KDT0 functional is constructed by mixing the short-range Fock exchange with the short-range thermal KDT16 exchange,
\bea
\mathcal{F}_{\text{xc}}^{\text{RS-KDT0}}[n,T] &=&
\mathcal{F}_{\text{xc}}^{\text{KDT16}}[n,T] \\
&+&
a\left(
\mathcal{F}_{\text{x}}^{\mathrm{SR,\mu F}}[n,T]
-
\mathcal{F}_{\text{x}}^{\mathrm{SR,\mu KDT16}}[n,T]
\right)
\,,
\nonumber
\eea
where $a=0.25$ denotes the fraction of short-range exact exchange. Details of the derivation are given in Ref.~\cite{Ammar_RSKDT0_2025}.

\subsection{Kubo-Greenwood Formalism}

The transport properties of warm dense matter are commonly studied using the Kubo--Greenwood formalism~\cite{kubo_1957,Greenwood_1958}. 
Within linear response theory, the electrical conductivity is an intrinsic material property that characterizes the response of an equilibrium system to a weak external perturbation and is independent of the strength of the applied field. 
The Kubo--Greenwood formulation is derived under the linear-response assumption, which allows transport coefficients to be expressed explicitly in terms of electronic eigen-states.

DFT provides access to the electronic structure for a given ionic configuration through the solution of the MKS equations. 
Using the resulting MKS eigen-states $\{\phi_n\}$ and eigenvalues $\{E_n\}$, the frequency-dependent Onsager coefficients can be written within the Kubo--Greenwood formalism~\cite{Suxing_CH_2014,Blanchet_Transport_abinit_2024} as
\begin{multline}
\mathcal{L}_{nm}(\omega) =
\frac{2\pi(-e)^{4-n-m}}{3\omega m_e^2 V}
\sum_{mn}
F_{mn} \, D_{mn} \\
\times
\left( \frac{E_m + E_n}{2} - h_e \right)^{n+m-2}
\delta(E_m - E_n - \hbar\omega),
\label{onsag_coef}
\end{multline}
where $e$ is the electron charge, $m_e$ is the electron mass, and $V$ is the system volume. 
The factor $F_{mn} = f_m - f_n$ denotes the difference between Fermi--Dirac occupation numbers at temperature $T$, $E_n$ is the KS eigenvalue of state $n$, and $h_e$ is the electronic enthalpy per particle. 
The quantity $D_{mn}$ represents the squared velocity dipole matrix elements, defined as
\begin{equation}
D_{mn} =
\left|
\left\langle \phi_m \middle| \nabla \middle| \phi_n \right\rangle
\right|^2 .
\end{equation}

The real part of the frequency-dependent electrical conductivity is directly obtained from the Onsager coefficient,
\begin{equation}
\sigma_1(\omega) = \mathcal{L}_{11}(\omega).
\end{equation}
The imaginary part of the conductivity is evaluated via the Kramers--Kronig relation as a principal value integral,
\begin{equation}
\sigma_2(\omega) =
-\frac{2}{\pi}
P \int
\frac{\omega \sigma_1(\omega')}{\omega'^{2} - \omega^2}
\, d\omega' .
\end{equation}
$P$ stands for the principal value of the integral. The real and imaginary parts of the dielectric function are then given by
\begin{align}
\epsilon_1(\omega) &= 1 - \frac{4\pi}{\omega}\sigma_2(\omega), \\
\epsilon_2(\omega) &= \frac{4\pi}{\omega}\sigma_1(\omega),
\end{align}
such that $\epsilon(\omega) = \epsilon_1(\omega) + i\epsilon_2(\omega)$.
From the complex dielectric function, the real and imaginary parts of the refractive index, $n(\omega)$ and $k(\omega)$, are obtained as
\begin{align}
n(\omega) &=
\sqrt{\frac{|\epsilon(\omega)| + \epsilon_1(\omega)}{2}}, \\
k(\omega) &=
\sqrt{\frac{|\epsilon(\omega)| - \epsilon_1(\omega)}{2}}.
\end{align}

\section{Computational details}
\label{sec:comp_details}
The AIMD simulations are performed using thermal DFT within plane wave  Vienna ab initio simulation package (VASP)~\cite{Kresse_VASP_PRB_1996,Kresse_VASP_CMC_1996}. Projector augmented wave (PAW)~~\cite{Blochl_PAW_PRB_1994,Kresse_PAW_PRB_1999} data sets are used with plane wave energy cutoff of 1400 eV. Hard PAWs with core radius of 0.8 \AA ~and 1.1 \AA ~for H and He respectively were selected. Thermal Tr$^2$SCANL meta-GGA XC functional~\cite{Karasiev_meta-GGA_2022} was used in these simulations.

We have performed simulations for H and He mixture with He fraction of $x_{\mathrm{He}}\equiv N_{\rm He}/(N_{\rm He}+N_{\rm H})\times 100\% = 11.3 \%$ in $NVT$ ensemble. 
Simulations were performed for 460 atoms (He$_{52}$H$_{408}$) for $T$ ranging from 3,000 K to 12,000 K, and for 230 atoms (He$_{26}$H$_{204}$) for $T$ ranging from 10,000 K to 12,000 K.  To accelerate the thermal equilibration, the system is assumed to be initially in the demixed state where He droplet is placed in the middle of simulation box with H in the boundaries. The initial positions of atoms do not affect the final state being mixed or demixed given the total time of the simulations is enough to reach equilibrium positions \cite{karasiev2026_H-He,Soubrian_Optical_2013}. At 10,000 K ($P=154.6~\mathrm{GPa}$) and 12,000 K ($P=174.8~\mathrm{GPa}$), the pressure differences between the 460- and 230-atom systems are approximately $0.13~\mathrm{GPa}$ and $0.053~\mathrm{GPa}$, respectively. The corresponding differences in specific internal energy are $0.089~\mathrm{kJ/g}$ and $0.062~\mathrm{kJ/g}$, respectively. 
Additionally, we preformed calculations for a small system size (He$_7$H$_{54}$) in which the system is always mixed. The largest pressure difference between 460~atom and 61~atom systems along the Hugoniot is $\approx 1$~GPa. The mixed system (61 atom) will be useful for investigating mixing and demixing state by radial distribution functional comparison.


The first Brillouin zone (BZ) is sampled using the Baldereschi mean-value point (BMVP)~\cite{PhysRevB.7.5212} at $(0.25, 0.25, 0.25)$ in $\bf k$-space for both 460-atom and 61-atom systems. The ionic temperature is controlled using a Nos\'e--Hoover thermostat. The ions are propagated according to the Newtonian equations of motion for approximately 10{,}000 MD steps for the large system and 20{,}000 MD steps for the small (mixed) system. The MD time step is chosen according to $dt \sim \rho^{-1/3} T^{-1/2}$,
with values ranging from 0.19 to 0.6~fs. The number of electronic bands is selected such that the highest energy band occupancy is approximately $2 \times 10^{-6}$.


The transport properties are calculated using the Kubo--Greenwood formalism as implemented in the VASP code. Between six and sixteen statistically independent snapshots are selected from the particle trajectories at each temperature along the Hugoniot. 
For single-point DFT calculations, a system containing 460 atom is used for $T \leq 8{,}000$~K and 230-atom for $T = 10{,}000$~K and $12{,}000$~K, while a small He$_7$H$_{54}$ mixed system is employed at all temperatures. 
We used the widely used PBE functional as well as the recently developed RS-KDT0 XC functional~\cite{Ammar_RSKDT0_2025}.

BZ is sampled using the BMV ${\bf k}$-point at $(0.25, 0.25, 0.25)$ for the large system, and a $3\times3\times3$ Monkhorst--Pack grid for the small mixed system. The number of electronic bands included in the transport calculations is chosen to be three times that used in the AIMD simulations. The Dirac delta function in Eq. (\ref{onsag_coef}) is broadened by a Gaussian with width $\Delta = 0.4$~eV for the 460- and 230-atom systems, and $\Delta = 0.6$~eV for the 61-atom system. Convergence tests were performed to ensure that the results presented here are converged with respect to $\bf k$-mesh and $\Delta$.

\section{Results and Discussions}
\label{results}
\subsection{EOS}
To determine the appropriate density along the princip\aae{al} Hugoniot at each temperature, we solved the Rankine–Hugoniot (RH) equation, which defines the thermodynamic relation between unshocked and shocked states:

\begin{equation}\label{eq_RHug}
    {\mathpzc E}_f - {\mathpzc E}_i + \frac{1}{2} \left( P_f + P_i \right)\left( \frac{1}{\rho_f} - \frac{1}{\rho_i} \right) = 0,
\end{equation}
where ${\mathpzc E}$, $P$, and $\rho$ denote the specific internal energy, pressure, and density, respectively. The subscripts ``$i$'' and ``$f$'' correspond to the unshocked (initial) and shocked (final) states.

The initial density was fixed at the average experimental value of $\rho_i = 0.274~\text{g/cm}^3$, which is pre-compressed by \aae{Diamond Anvil Cell} (DAC) in experiments~\cite{Byrgoo_H-He_2021}. We obtained an initial specific internal energy ${\mathpzc E}_i = -212.2$ kJ/g 
and pressure of $P_i \approx 3.1~\text{GPa}$, in good agreement with the pre-compression pressure of 4~GPa used in the experiment. At each temperature, we selected two to three trial densities and performed AIMD simulations to compute the corresponding specific internal energy and pressure. We then interpolated these results to identify the shocked density that satisfies the RH condition.


%
 \begin{figure}[ht]
{\includegraphics[width= 0.5\textwidth]{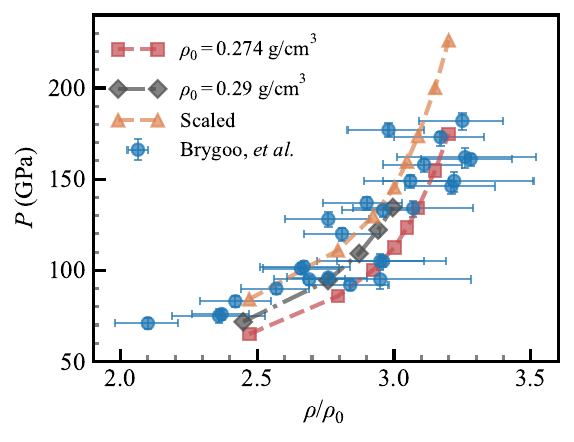}}
\caption{Shock pressure as a function  of density along the principal Hugoniot of H-He mixture (11.3\% He). Results of AIMD simulations
with meta-GGA XC Tr$^2$SCANL (\aae{red squares for $\rho_i = 0.274$ g/cm$^3$, dark gray diamond for $\rho_i = 0.29$ g/cm$^3$, and dark yellow triangles for scaled $\rho_i = 0.274$ g/cm$^3$ }) are compared to experimental values \cite{Byrgoo_H-He_2021} (blue circles).
}
\label{fig:R_vs_P}
\end{figure}

\begin{figure}[ht]
{\includegraphics[width= 0.5\textwidth]{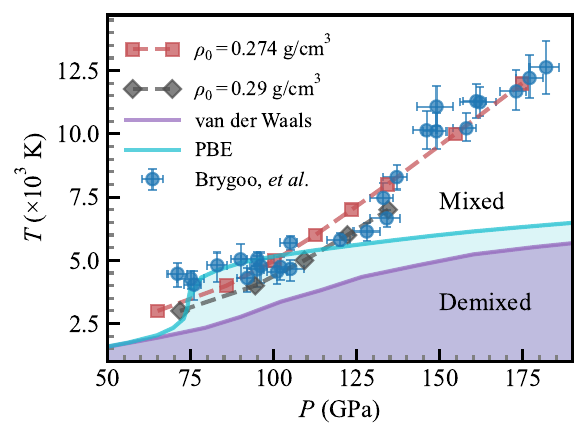}}
\caption{Temperature as a function of pressure along the principal Hugoniot of H-He mixture (11.3\% He). The Tr$^2$SCANL AIMD results (\aae{red squares for $\rho_i = 0.274$ g/cm$^3$, and dark gray diamond for $\rho_i = 0.29$ g/cm$^3$}) are compared to experimental values \cite{Byrgoo_H-He_2021} (blue circles).
The immiscibility boundaries previously calculated using PBE~\cite{Morales_H-He_2013} and van der Waals~\cite{schottler_H-He_2018} XC functionals for H-He mixture with 8 $\%$ (atomic fraction) He are shown in light blue and purple, respectively.  
}

\label{fig:P_vs_T}
\end{figure}

\aae{Figure~\ref{fig:R_vs_P} presents the pressure of the H–He mixture (11.3$\%$ He) versus \aae{compression ratio $\rho/\rho_i$} along the principal Hugoniot. The results using the Tr$^2$SCANL meta-GGA thermal XC functional~\cite{Karasiev_meta-GGA_2022} show overall the same trend as the experimental data, within uncertainties, except at $\rho/\rho_i = 0.47$ ($\rho = $ 0.677 g/cm$^3$) and ($\rho = $0.835 g/cm$^3$) corresponding to 3,000 K (65.0 GPa) and 7,000 K (123.4 GPa) respectively. We note that 3,000 K is lower than the recorded experimental values for temperature around 4,500 K (see Fig. \ref{fig:P_vs_T}) which could have an effect on the estimated pressure. Additionally, the accuracy of the estimation of the initial unshocked system density $\rho_i$, and related energy ${\mathpzc E}_i$, can affect the Hugoniot curve. We summarize the results of the simulations along the Hugoniot curve with initial density $\rho_i = 0.274,\mathrm{g/cm^3}$ in Table~\ref{table:1}.} 

\aae{We performed an additional set of simulations to investigate changes in the Hugoniot curve in the low-density region when the initial density is increased to $\rho_i = 0.29$ g/cm$^3$, which is higher than the value reported in the experiment. For $\rho_i = 0.29$ g/cm$^3$, we obtained an initial energy ${\mathpzc E}_i = -212.0$ kJ/g and an initial pressure $P_i = 3.6$ GPa, which is closer to the experimental value of 4 GPa. The resulting Hugoniot curve shows better agreement with the experimental data in the low-pressure regime, indicating that the deviation found for $\rho_i = 0.274$ g/cm$^3$ may arise from the estimate of $\rho_i$.}

\aae{To further examine the uncertainty in $\rho_i$, we note that Eq.~(\ref{eq_RHug}) is invariant under the scaling transformation $\rho \rightarrow r\rho$, $P \rightarrow \pi P$, and ${\mathpzc E} \rightarrow (\pi/r){\mathpzc E}$. If $P$ is scaled so that $P_i$ matches the experimental value of $4$ GPa, then $\pi = 4/3.1 \approx 1.29$. Thus, for the scaled Hugoniot for $\rho_i =0.274$ g/cm$^3$, $P(\rho/\rho_i) \rightarrow \pi P(\rho/\rho_i)$ with compression is invariant under transformation. In Figure~\ref{fig:R_vs_P}, the scaled $P(\rho/\rho_i)$ shows the higher pressure values compared to other Huogoniot curves within experimental values in low pressure regime, although it overestimate the pressure for higher compression values. If one chooses ${\mathpzc E} \rightarrow {\mathpzc E}$, this implies $\pi/r = 1$, or equivalently $r = \pi \approx 1.3$ resulting in $r\rho_i \approx 0.354$ g/cm$^3$.}

\begin{table}[ht]
\centering
\caption{\aae{EOS results from AIMD simulations using the Tr$^2$SCANL XC functional for the Hugoniot curve with an initial density of $\rho_i = 0.274$ g/cm$^3$.}}
\begin{tabular}{c c c c c c} 
 \hline\hline 
 \\T (K)& ~~$\rho$ (g/cm$^3$)~~ & ~~$\rho/\rho_i$~~& ~~$P$ (GPa)~~& ~~$u$ (kJ/g)~~\\ [0.5ex] 
 \hline
 3,000 & ~~0.677~~ & ~~2.47~~ &~~65.0~~ & ~~-139.5~~\\
 4,000 & ~0.766 & 2.79 &~85.8 & ~-107.7\\
 5,000 & ~0.801 & 2.92 &~100.2 & ~-88.1 \\ 
 6,000 & ~0.822 & 3.00 &~112.5 & ~-72.2 \\ 
 7,000 & ~0.835 & 3.05 &~123.3 & ~-57.2 \\ 
 8,000 & ~0.846 & 3.08 &~134.2 & ~-43.1 \\ 
 10,000 & ~0.863 & 3.15 &~154.6 & ~-15.6 \\ 
 12,000 & ~0.876 & 3.2 &~174.8 & ~11.0 \\ 
 [1ex]
 \hline \hline
\end{tabular}
\label{table:1}
\end{table}
Figure~\ref{fig:P_vs_T} displays the corresponding temperature as a function of pressure along the Hugoniot. The theoretical predictions reproduce the overall increasing trend of the shock temperature observed experimentally. \aae{However, the temperature discontinuity between 6000 and 8000 K (110 and 150 GPa) reported in the experiment is not observed in our simulations with $\rho_i = 0.274$ g/cm$^3$. In contrast, the Hugoniot curve obtained with $\rho_i = 0.29$ g/cm$^3$ shows good agreement with the experimental results in this regime, suggesting that the discrepancy may arise from the estimate of the initial density.} Additionally, the absence of this feature in our simulations may indicate that additional physical effects are present in the experiment but are not captured within the present AIMD framework. For example, shock compression can induce species separation at the shock front~\cite{Zhang_2020}. Whether such effects contribute to the observed temperature discontinuity remains an open question. 

The theoretical predictions of the van der Waals~\cite{schottler_H-He_2018} and PBE~\cite{Morales_H-He_2013} immiscibility boundaries are shown in purple and cyan respectively. These immiscibility boundaries are calculated using thermodynamics relations with calculating the non-ideal entropy of mixing using coupling constant and thermodynamics integrations. According to van der Waals calculations, the system is always mixed during the shock compression. However, for the PBE the system is in the demixed state in the interval of  3,500 K to 5,000 K. 

\subsection{Radial distribution functions}

The demixing state of the system can be qualitatively examined from the trajectories and spatial distributions of ions within the simulation cell~\cite{Lorenzen_2011,Militzer_H-He_2013,Hamel_H-He_2011,karasiev2026_H-He}. Radial distribution functions (RDFs) provide clear structural signatures of demixing. In a demixed state, like atoms tend to cluster together while unlike atoms are more spatially separated compared to a mixed state. This behavior is reflected in the fact that $g_{\text{H-He}}(r)$ decreases at low distances compared to homogeneous (mixed) distributions of ions. 

\begin{figure*}[ht]
{\includegraphics[width= 0.85\textwidth]{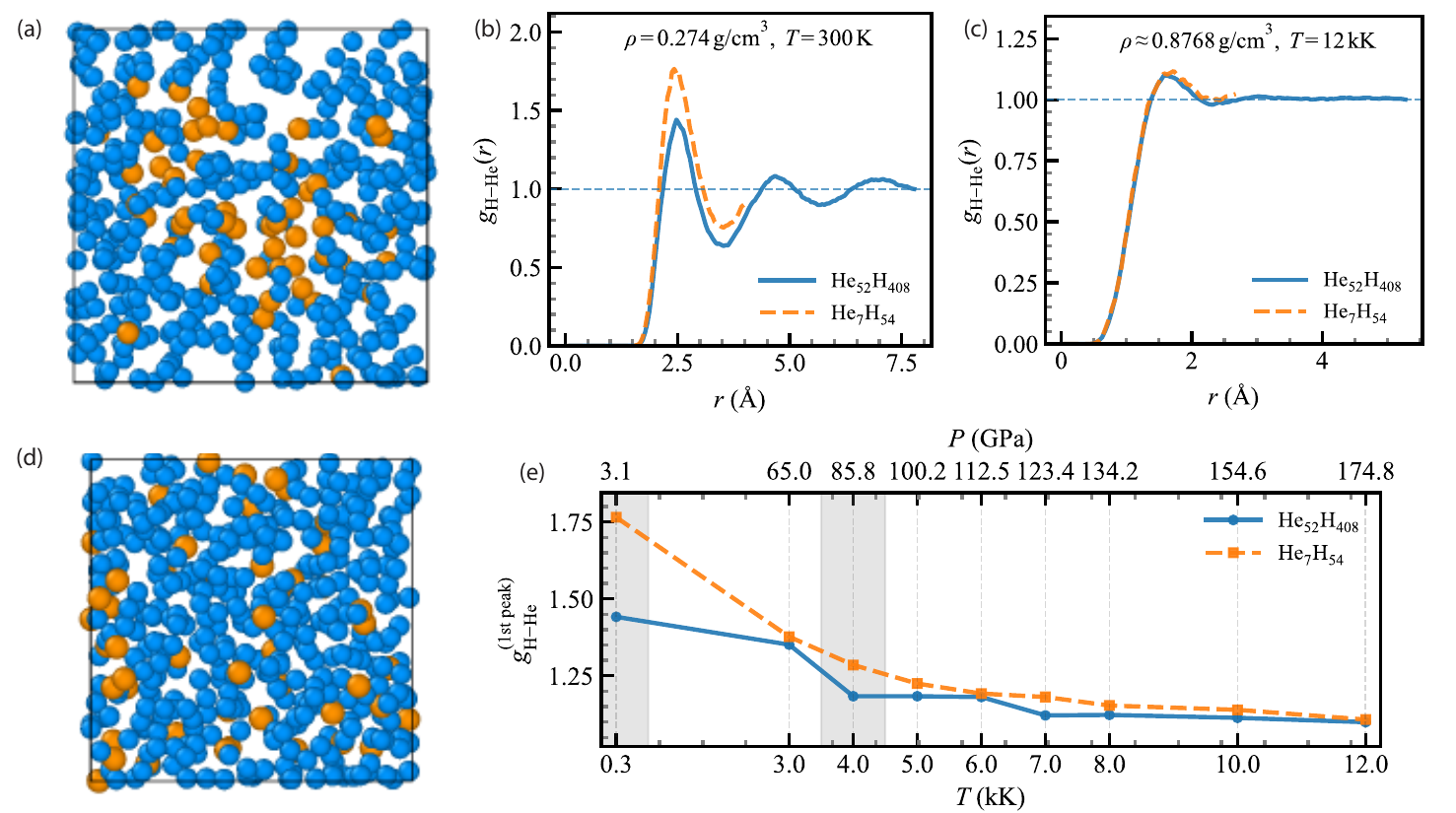}} 
\caption{\aae{Snapshots of mixed and demixed large size systems, and H–He radial distribution functions for large and small system sizes. Panel (a) shows a snapshot of H–He at the initial conditions of the unshocked mixture, where the system is demixed, and panel (b) shows the corresponding H–He radial distribution function. Panel (c) shows the H–He radial distribution function for the shocked mixture at $T = 12{,}000$ K, and panel (d) shows a snapshot of H–He at $T = 12{,}000$ K, where the system is mixed.}
Panel (e) shows the first-peak height of the H–He RDF along the Hugoniot for both large and small system sizes.
}
\label{fig:g_of_r}
\end{figure*}

To compare with fully mixed state, we performed AIMD simulations for a small system containing 54 hydrogen and 7 helium atoms along the principal Hugoniot. The small system size ensures that the system is always mixed. Figures~\ref{fig:g_of_r} (a)  and (b)  show the H-He radial distribution function, $g_{\text{H}\text{--}\text{He}}(r)$, for two conditions: the initial state at $T = 300$~K and the highest temperature $T = 12,000$~K along the Hugoniot. At the initial state, the first-peak height of the 460-atom system is lower than that of the 61-atom system (representing the fully mixed case) with $\Delta g_{\mathrm{H\text{--}He}}^{(1\mathrm{st}\,\mathrm{peak})} = 0.32$ (approximately 22\%), indicating a degree of demixing. At $T = 12,000$~K, however, the peak heights of both systems coincide, which is consistent with a mixed state.

To obtain a qualitative measure of the mixing behavior along the Hugoniot, we plot the height of the first RDF peak in Fig.~\ref{fig:g_of_r} (c). The difference in the first-peak height of $g_{\text{H}\text{--}\text{He}}(r)$ between the two systems do not decrease monotonically with increase of temperature and pressure along the Hugoniot. Instead, larger differences are observed at $T = 4,000$~K with $\Delta g_{\mathrm{H\text{--}He}}^{(1\mathrm{st}\,\mathrm{peak})} = 0.10$ (approximately 9\%) compared to the value at $T = 3{,}000$~K, where $\Delta g_{\mathrm{H\text{--}He}}^{(1\mathrm{st}\,\mathrm{peak})} = 0.03$ (approximately 2\%), followed by a reduction at higher temperatures, 
reaching minimal differences at $T = 10,000$~K and $12,000$~K. Although a slight increase is observed around 7000~K, with 
$\Delta g_{\mathrm{H\text{--}He}}^{(1\mathrm{st}\,\mathrm{peak})} = 0.06$ 
(approximately 5\%), this percentage change is small to classify the mixture as demixed at this temperature. Accurate predictions of the demixing boundaries require observation of the behavior of $g_{\mathrm{H\text{--}He}}^{(1\mathrm{st}\mathrm{peak})}$ with increasing $T$ along selected isobars, until it reaches the maximum value due to mixing and starts to decrease due to thermal expansion (see details in \cite{karasiev2026_H-He}). These results suggest that the degree of mixing varies along the Hugoniot, with partial demixing occurring around $4,000$~K. 
This agrees with PBE immiscibility boundary~\cite{Morales_H-He_2013}. 

\begin{figure}[ht]
{\includegraphics[width= 0.5\textwidth]{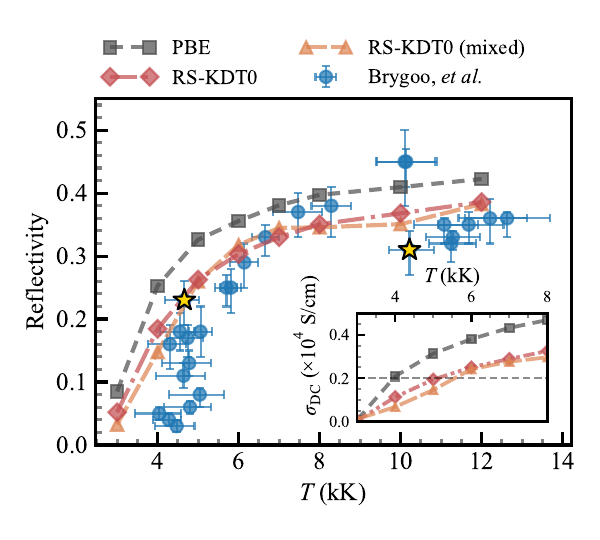}}
\caption{Reflectivity for a shock-compressed H–He mixture along the principal Hugoniot \aae{for initial density $\rho_i = 0.274$ g/cm$^3$}. Experimental values~\cite{Byrgoo_H-He_2021} are shown in blue, with gold stars indicating the reflectivity jump points. Results for the large system size using PBE and RS-KDT0 are shown in green and red, respectively. Results for the small (mixed) system are shown in orange. The inset panel shows the DC conductivity of the mixture. The horizontal line corresponds to Mott’s criterion for the minimum metallic conductivity, $2 \times 10^{3}~\mathrm{S/cm}$. } 

\label{fig:reflectivity}
\end{figure}
\subsection{Optical Properties}

Following the AIMD simulations, we selected several independent snapshots of particle trajectories at each temperature and density along the Hugoniot curve \aae{for initial density $\rho_i = 0.274$ g/cm$^3$} to compute the optical conductivity and reflectivity. For each snapshot, we performed single-point DFT calculations using the PBE and RS-KDT0 XC functionals. The reflectivity is then calculated for monochromatic light with wave length 532 nm incident on the shock front, the interface between shocked and unshocked regions, assuming the unshocked medium has a real refractive index. The reflectivity is given by:
\begin{equation}
    R(\omega) = \frac{[n(\omega) - n_0]^2 + k(\omega)^2}{[n(\omega) + n_0]^2 + k(\omega)^2},
\end{equation}
where $n_0$ is the refractive index of the unshocked H--He mixture \aae{with density $\rho_i = 0.274$ g/cm$^3$} which is equal to 1.34 and 1.28 when calculated with PBE and RS-KDT0 XC functionals respectively. 

Figure~\ref{fig:reflectivity} compares the calculated reflectivity with experimental data along the shock Hugoniot. The gold stars denote reflectivity jump/drop corresponding to the demixing and mixing boundaries observed in recent shock-compression experiments~\cite{Byrgoo_H-He_2021}. 
First, the reflectivity values calculated with PBE are generally overestimated with respect to experimental values, except  
for higher values around~ 10,000 K. This overestimation of reflectivity can be linked to the underestimations of band gap resulting in higher conductivity values. On the other hand, the calculations with RS-KDT0 provide better agreement with experiment across the full temperature range (except the values around 10,000 K) within the error bars, capturing the trend in reflectivity growth with temperature. Notably, while all theoretical curves show a monotonic increase of reflectivity, they do not reproduce the discrete experimental drop at elevated temperature associated with mixing.

\aae{Additionally, we performed reflectivity calculations using the RS-KDT0 XC functional for the 61-atom system as a representative mixed-system configuration. The reflectivity obtained for the 61-atom system agrees well with that of the large 460-atom system over the temperature range from 5{,}000 to 12{,}000~K, with a maximum deviation of approximately 5\% at $T = 10{,}000$~K. At lower temperature, a larger deviation of about 20\% (with reflectivity of 460-atom and 61-atom systems equal to 0.184 and 0.148 respectively)  is observed at $T = 4{,}000$~K, near the first experimentally observed jump in reflectivity. The DC conductivity, shown in the inset of Fig.~\ref{fig:reflectivity}, is lower for the 61-atom (mixed) system than for the 460-atom system in the temperature range $3500$--$5500$~K. 
The IMT is identified using Mott's criterion for the minimum metallic conductivity, $2000_{-1000}^{+3000}~\mathrm{S/cm}$~\cite{karasiev2026_H-He,Lorenzen_2011}. 
The large system undergoes the IMT at 5,000 K, that coincides with the jump of reflectivity observed in the experiment, marked by gold star in Fig. \ref{fig:reflectivity}.
It has been suggested that metallization enhances demixing in H--He mixtures~\cite{karasiev2026_H-He,Lorenzen_2011}, which may explain the demixed state observed around $4000$~K in our simulations. The difference in reflectivity at $4000$~K may therefore be attributed to the system being in a demixed state, as suggested by Soubiran \emph{et al.}~\cite{Soubrian_Optical_2013}.}
 
 \aae{Moreover, our result show that the calculated reflectivity of fully mixed system agrees the experimentally measured reflectivity in the temperature range of 5{,}000~K – 8{,}000~K (100.2 GPa - 134.2 GPa), where the mixture was inferred to be demixed from experiments. We note that one of the arguments for demixing presented in Ref.~\cite{Byrgoo_H-He_2021} is that the reflectivity of an H--He mixture is lower than that of pure hydrogen under the same pressure and temperature conditions. They adopted a simplified model in which the reflectivity of a homogeneous H--He mixture at a given temperature is mapped onto that of pure hydrogen through the introduction of an effective hydrogen density, $\rho_{\mathrm{H_{eff}}}$. This effective density is defined in terms of the mixture density $\rho_{\mathrm{mixture}}$ as
 $\rho_{\mathrm{H_{eff}}}=x_{\mathrm{H}}/(1/\rho_{\mathrm{mixture}}-x_{\mathrm{He}} a^{3})$,
where $a = 1.05\,\text{\AA}$, and $x_{\mathrm{H}}$ and $x_{\mathrm{He}}$ denote the hydrogen and helium fractions, respectively. In the temperature range of $5{,}000$--$8{,}000$~K, this model underestimates the measured reflectivity of the $11\%$ H--He mixture, which instead appears to follow the behavior of pure hydrogen. They associate the reflectivity increase in this range as demixing signature. Although this model appears to reproduce the experimental reflectivity for the $33\%$ H--He mixture, this composition has been reported to remain in the mixed state. Additionally, we note that in the work of Soubiran \emph{et al.}~\cite{Soubrian_Optical_2013}, the degree of mixing in a system with a 50\% helium molar fraction was shown to significantly affect the reflectivity. However, lower helium concentrations were not investigated, leaving open the question of whether the expected reflectivity contrast between mixed and demixed states remains sufficiently pronounced at small helium fractions. }
\aae{Furthermore, the shock may kinetically drive hydrogen ahead of helium at the shock front, such that VISAR could possibly probe a shock-induced separation. If this occurs in the experiment~\cite{Byrgoo_H-He_2021}, then the observed hydrogen-helium separation would not necessarily correspond to thermodynamically driven demixing.}


\section{Summary}
\label{summary}
In this work, we investigated the properties of an H--He mixture with 11.3\% He atomic fraction along the principal Hugoniot, motivated by recent shock-driven experiments. The calculations were performed using thermal XC functionals within DFT, combined with the Kubo--Greenwood formalism for transport properties. The equation of state was obtained using the thermal meta-GGA Tr$^2$SCANL functional. 
The predicted Hugoniot curve \aae{for $\rho_i = 0.274$ g/cm$^3$} shows overall good agreement with the experimental data, remaining within the experimental error bars except at $\rho/\rho_i = 2.47$ ($\rho = 0.677~\mathrm{g/cm^{3}}$), corresponding to a temperature of approximately 3,000~K. \aae{This discrepancy appears to be sensitive to the choice of initial density: when the initial density is increased to $\rho_i = 0.29$ g/cm$^3$, the resulting Hugoniot curve shows improved agreement with the experimental data in the low-pressure regime (70--80 GPa). This suggests that the deviation observed for $\rho_i = 0.274$ g/cm$^3$ may arise from uncertainties in the initial-state density and the corresponding initial energy ${\mathpzc E}_i$, both of which directly affect the predicted low-pressure shock data.}

The predicted pressure--temperature relation generally follows the trend of the experimental measurements; however, the kink observed experimentally around 6,000--8,000~K is absent in our simulations, leading to a systematic underestimation of the pressure in this temperature range. \aae{This discrepancy may be related to the choice of initial density, as the Hugoniot curve computed with an initial density of $\rho_i = 0.29$ g/cm$^3$ shows better agreement with the experimental data in the 6,000--7,000~K regime.}

Demixing behavior was further examined directly from the simulations by analyzing the first peak of the H--He radial distribution functions for two system sizes: a 460-atom cell and a small 61-atom mixed system. Pronounced deviations were observed at 300~K and 4,000~K, consistent with the demixing trends predicted by the PBE immiscibility curve, possibly higher by 500 K.

Optical reflectivity was calculated using the thermal range separated RS-KDT0 hybrid XC functional, which provides improved band-gap predictions at elevated temperatures. The predicted reflectivities for both system sizes are in overall agreement with experimental values within the reported uncertainties, except at $T = 10,000$~K where no drop in reflectivity was observed in the simulations, in contrast to experimental reports. The close agreement between the two system sizes---particularly in the 5000--8000~K temperature range, where experiments suggest immiscibility---raises questions about the sensitivity of reflectivity as a diagnostic for demixing at low helium concentrations.

These findings suggest that further investigation of reflectivity signatures across a broader range of helium fractions is necessary, especially given that previous studies such as Ref.~\cite{Soubrian_Optical_2013} focused primarily on 50\%/50\% H-He mixtures. Better understanding of what can possibly occur in experiments (e.g. shock-induced species separation, the H$_2$ subsystem dissociation, the insulator-to-metal transition) would further clarifying the true signature of thermodynamically-driven H-He demixing. In addition, identifying alternative experimentally accessible signatures of demixing in H--He mixtures may be essential for reliably probing phase separation in the low-helium regime. 


\begin{acknowledgments}
\aae{We acknowledge the anonymous referee for the helpful discussion regarding the scaling of the Hugoniot equation. }This work is supported by the Department of Energy [National Nuclear Security Administration] University of Rochester “National Inertial Confinement Fusion Program” under Award Number DE-NA0004144 and U.S. National Science Foundation PHY Grant No. 2020249.
 
This report was prepared as an account of work sponsored by an agency of the U.S. Government. Neither the U.S. Government nor any agency thereof, nor any of their employees, makes any warranty, express or implied, for the accuracy, completeness, or usefulness of any information, apparatus, product, or process disclosed, or represents that its use would not infringe privately owned rights. Reference herein to any specific commercial product, process, or service by trade name, trademark, manufacturer, or otherwise does not neces- sarily constitute or imply its endorsement, recommendation, or favoring by the U.S. Government or any agency thereof. The views and opinions of authors expressed herein do not necessarily state or reflect those of the U.S. Government or any agency thereof.
\end{acknowledgments}


%
\end{document}